\documentclass[aps,prl,twocolumn,showpacs,amsmath,amssymb,superscriptaddress,10pt,floatfix]{revtex4-2}
\usepackage{graphicx} 
\usepackage{orcidlink}
\usepackage{qcircuit}
\usepackage[T1]{fontenc}

\begin{document}

\title{Noise-resilient and Scalable Quantum Error Correction for Nuclear Spin Qubits in Silicon with Electron Shuttling}
\author{Wayne M. \surname{Witzel}\orcidlink{0000-0002-9082-8076}}
\affiliation{Center for Computing Research, Sandia National Laboratories, 
Albuquerque, New Mexico 87185 USA}
\author{Matthew D. \surname{Grace}\orcidlink{0000-0001-5842-7347}}
\affiliation{Center for Computing Research, Sandia National Laboratories, 
Albuquerque, New Mexico 87185 USA}
\author{Dwight R. \surname{Luhman}\orcidlink{0000-0001-8045-1132}}
\affiliation{Sandia National Laboratories, Albuquerque, New Mexico 87185 USA}

\date{\today}

\begin{abstract}
Nuclear spin qubits in silicon are well-isolated from their environment.  Consequently, they have very long lifetimes and low sensitivity to noise, but this also suggests that control and measurement is challenging.  We introduce electron pair interferometry (EPI), a protocol to overcome this challenge and maintain robustness to noise. EPI is implemented using an array of quantum dots with isoelectronic nuclear spin qubits located in the dots. A pair of electrons are initialized into a singlet ground state, split apart, and shuttled to the dots containing nuclear spin qubits. We show it is possible to coherently transfer the parity of the nuclei onto the measurable state of singlet/triplet-encoded electrons.  Global nuclear magnetic resonance (NMR) can be used to change bases and implement dynamical decoupling (DD).  Combined with selective hyperfine-induced $Z_{\pi}$ rotations, our gate set is complete for universal quantum computation, tailored to Calderbank-Shor-Steane (CSS) quantum error correction, and robust to noise.  We discuss very low sensitivity to charge noise and study the sensitivity to both DC and AC magnetic field inhomogeneity which depends strongly on their relative strengths.
\end{abstract}
\maketitle

\section{Introduction}

Promising qubit platforms simultaneously balance the demands of long coherence times, high fidelity entanglement operations and scalability. Nuclear spin qubits in silicon have demonstrated extraordinarily long coherence times and high fidelity control through nuclear magnetic resonance (NMR)\cite{Muhonen2014, Maurer2012, Saeedi2013}. Direct entanglement between nuclear spins qubits is challenging due to the very weak dipolar interaction. A natural solution to overcome this is to use the hyperfine interaction between an electron and nucleus, as in donor qubit systems, together with the exchange interaction between electrons on neighboring nuclei\cite{Kane1998}. In this scenario the electron spins act as ancilla for entanglement between the nuclei. The implementation of this scheme requires control over the hyperfine and exchange interactions. Donor based systems require close spatial proximity between the donors such that the wavefunction overlap of the bound electrons results in a strong exchange interaction. Recent experiments in silicon devices have demonstrated this manner of entanglement between $^{31}$P nuclei in stochastically implanted donors \cite{Stemp2025} and deterministically placed donors \cite{Thorvaldson2025,Edlbauer2025}. Scaling these approaches to large arrays of reliably interacting nuclear spin qubits remains an outstanding engineering challenge.

Quantum dots are a fundamental building block of spin qubit architectures and efforts to scale silicon-based quantum dot devices has accelerated in recent years with demonstrations of linear quantum dot arrays \cite{Struck2024, Xue2024, Philips2022, Madzik2025, Nickl2026} and, remarkably, two dimensional systems \cite{Wang2D2024, HanLim2025, Unseld2025, HRLpreprint}.
Coherent shuttling of electrons across quantum dot arrays in silicon has also become routine\cite{Yoneda2021, Seidler2022, Struck2024, Xue2024, Volmer2024, Losert2024, Foster2025, Langheinrich2025, Lin2025preprint, Volmer2025preprint}. Hyperfine coupling between an electron in a quantum dot and an embedded single nuclear spin of an isoelectronic species has even been demonstrated for $^{29}$Si\cite{Hensen2020-bf} and $^{73}$Ge\cite{SteinackerArxiv}.

Collectively, the advancements in silicon quantum dot devices highlight the increasing feasibility of an elegant approach to a scalable nuclear spin based quantum architecture. In this scenario, the data qubits are NMR-controlled nuclear spins of single isoelectronic atoms located in silicon quantum dots and ancilla electrons are shuttled between quantum dot sites. Long-range entanglement of data qubits is achieved via the sequential hyperfine interaction of ancilla electrons with data qubits, specifically hyperfine CPhase gates~\footnote{referred to as electron-nuclear controlled-phase or e-n-CPhase gates in Ref.~\cite{Witzel2022}}~\cite{Witzel2022}. Precise control of entanglement in this scheme is achieved through the duration of the hyperfine interaction, which is easily set by shuttling the electron on and off the quantum dots containing target data qubits (or avoiding them by simply shuttling through quickly). The confinement potential of a quantum dot can be tuned to optimize the hyperfine interaction to accommodate the specific location of the nuclear spin qubit within the dot, easing the placement requirements of the isoelectronic species. An analysis of error channels of the hyperfine CPhase operation in the context of isoelectronic Sn nuclear spin qubits in silicon quantum dots has shown the potential for very low error rates when the isotopic background of $^{29}$Si nuclear spins is reduced to levels of $\lesssim 50$ ppm.
Recent advancements in isotopically enriched silicon layers have reached single digit ppm concentrations of $^{29}$Si \cite{Acharya2024, Lim2025}, further elevating the appeal of this approach.

Here, we advance this architecture by introducing the concept of electron pair interferometry as a foundation for a novel approach to quantum error correction. The approach exploits the precision of global NMR and the flexibility of electron shuttling and hyperfine CPhase operations at the physical qubit layer to create a noise-resilient approach for universal quantum computation using nuclear spins.

\section{Electron pair interferometry}

Interferometry is a measurement technique based on splitting a wave along multiple paths, recombining the wave, and measuring the interference effects.  Electron pair interferometry (EPI) applies this concept to a pair of net-unpolarized electrons traveling along two distinct paths, recombining them, and measuring the interference effects manifest in the spin state probabilities. In EPI the ground state of the electron pair is the singlet state, $\lvert S \rangle = \left(\lvert \uparrow \downarrow \rangle - \lvert \downarrow \uparrow \rangle \right)/\sqrt{2}$, prepared in an electrostatically defined quantum dot, where the exchange interaction dominates over Zeeman and other interactions.  By splitting such a pair and adiabatically shuttling each electron along a distinct path, the wavefunction becomes a superposition in which the individual up and down spins follow distinct paths in both combinations.  Using $A$ and $B$ to label the paths, the wavefunction becomes $\lvert \psi(t) \rangle = \left(\lvert \uparrow \rangle_A \lvert \downarrow \rangle_B - e^{-i \phi(t)} \lvert \downarrow \rangle_A \lvert \uparrow \rangle_B \right)/\sqrt{2}$ where $\phi(t)$ is determined by the difference in magnetic fields, integrated over time, of the two paths.  Measuring singlet versus triplet at time $t_{\rm M}$ will ideally result in a singlet return probability of $P_S = \| \langle S \vert \psi(t_{\rm M}) \rangle \|^2 = \cos^2{\left(\phi(t_{\rm M})/2\right)}$ which can be measured using a spin-to-charge conversion process such as Pauli spin blockade.

The parity of a set of spin-$1/2$ nuclei can be determined using EPI by visiting one subset of nuclear spins along path $A$ and visiting a complementary subset along path $B$. The spin states of the visited nuclei will determine the resulting value of $\phi(t_{\rm M})$. EPI is only sensitive to the magnitude of the magnetic differences and not the sign of the difference since $P_S = \cos^2{\left(\phi(t_{\rm M})/2\right)}$ is an even function of $\phi(t_{\rm M})$. For this reason, it is most straightforward to measure the parity of an even number of spin-$1/2$ nuclei.

A parity check of an even number of nuclei, or data qubits, is achieved by divvying the nuclei evenly among the two paths and timing interactions with each nuclear spin so they contribute approximately $\pm \pi/2$ to $\phi$ with the sign dependent upon the nuclear spin polarization as well as which path visits that nucleus. This nuclear-dependent effect is implemented with a hyperfine CPhase operation~\cite{Witzel2022} which is performed by shuttling an electron into the quantum dot containing the nuclear spin qubit, waiting for an appropriate amount of time, and then shuttling away. The overall effect of this electro-nuclear interaction, assuming adiabaticity, is a CPhase operation between the two spin qubits up to local $Z$ rotations.
If all of the nuclei have the same polarization, so that both paths give equal but opposite contributions to $\phi$, then $\phi=0$ and the singlet state will be measured indicating an even parity. If one nucleus has the opposite polarization, $\phi=\pm\pi$ and the triplet state will be measured indicating an odd parity. With each nuclear polarization flip from that point, the parity flips and so does the measurement outcome.

A 2-parity measurement is depicted as
\begin{equation}
\label{eq:2parity}
\Qcircuit @C=1em @R=0.5em {
\lstick{} & \ctrl{2} & \gate{Z_{a_1}} & \qw & \qw & 
\gate{Z_{c_1}} & \multimeasureD{1}{\text{S/T}} \\
\lstick{} & \qw & \qw & \ctrl{2} & \gate{Z_{a_2}} &
\gate{Z_{c_2}} & \ghost{\text{S/T}}
\inputgroupv{1}{2}{.8em}{.8em}{\lvert S \rangle}\\
& \control \qw & \gate{Z_{b_1}} & \qw & \qw & \qw & \qw \\
& \qw & \qw & \control \qw & \gate{Z_{b_2}} & \qw & \qw
\gategroup{1}{2}{3}{3}{.7em}{--}
\gategroup{2}{4}{4}{5}{.7em}{--}
}
\end{equation}
in quantum circuit form where the top two wires represent the pair of electrons and the bottom two wires represent the two nuclear qubits.  Dashed boxes are drawn around the hyperfine-CPhase gates which produce local $Z$ rotations (with $a_1$, $a_2$, $b_1$, and $b_2$ rotation angles) on each spin as well as the controlled-$Z$ operation.  The angles $c_1$ and $c_2$ account for any additional $Z$-rotations on the electrons during shuttling.

The net effect of $\phi$ prior to measurement is
\begin{equation}
\phi = \phi_{\rm N} + a_1 + c_1 - a_2 - c_2,
\end{equation}
where $\phi_{\rm N}$ depends on the nuclear states (via the CPhase operations); $\phi_{\rm N} = 0$ when both nuclear polarizations are the same and $\phi_{\rm N} = \pm \pi$ when they differ.
In the ideal scenario $a_1 + c_1 = a_2 + c_2$ and $\phi$ is determined solely by $\phi_{\rm N}$ with even parity (same polarization) resulting in a singlet measurement outcome and odd parity (differing polarization) resulting in a triplet measurement outcome.

The local $Z$ rotation angles, $a_1$, $a_2$, $b_1$, and $b_2$, are determined by the magnetic field and hyperfine interactions strengths.  
For convenience, we work in the rotating frame governed by the Zeeman interaction with each qubit.
In this frame of reference and in a sufficiently large and uniform $B$-field such that the Zeeman difference is large relative to the hyperfine interaction, $a_1 \approx a_2 \approx b_1 \approx b_2 \approx -\pi/2$ such that the overall phase incurred
on each electron and on each nucleus when including effects of the CPhase gates
is approximately $\pm \pi/2$ with the sign depending on the state of the nucleus/electron with which it interacts.  
The precise value of these constants is determined by the magnetic field and the respective hyperfine interactions and will necessarily differ when the hyperfine interactions differ, even in the ideal case.
The impact of any relative difference between the nuclear spin rotations, $b_1$ and $b_2$, that is consistent in time (i.e., variability in space but not time) can be canceled by dynamical decoupling (DD) (discussed in the next section).
The error due to $a_1 \neq a_2$ could be accepted or mitigated by compensating with $c_1$ and $c_2$ 
to the extent they are controllable.
The shuttle-induced electron rotations $c_1$ and $c_2$ can also be minimized by having one electron follow the other during a portion of their journeys, a different form of DD.

As a concrete example of the effect of $a_1 \neq a_2$, if it is not compensated by $c_1$ and $c_2$, consider $^{119}$Sn qubits which are expected to have a relatively large hyperfine interaction of hundreds of kHz ~\cite{Witzel2022}. The nuclear gyromagnetic ratio is 16~MHz/T versus 28~GHz/T of an electron in silicon.
If the hyperfine interactions of the two qubits are respectively 100~kHz and 400~kHz and the $B$-field is one millitesla, 
$a_1-a_2$ will contribute about $0.017$ radians to $\phi$ resulting in a measurement outcome error on its own of $1-\cos^2{\left(0.017/2\right)} \approx 3 \times 10^{-4}$.  While relatively small in this instance, keep in mind that such errors can add coherently when measuring the parity of a larger number of nuclear spins, linearly in $\phi$ and quadratically in the error probability.

Many of the well-studied CSS codes, such as surface and color codes, require parity checks solely on even numbers of data qubits for syndrome extraction and are well suited to the EPI approach.  However, logical qubit preparation and measurement techniques often require measurements of individual data qubits, or an odd number of nuclei.
It is only possible to measure the parity of an odd number of nuclear spins using EPI if singlet/triplet rotations may be induced controllably to determine $c_1 - c_2$.
This could be effected, for example, with a micromagnet or intrinsic spin-orbit interactions~\cite{Jock2018}.
In general, minimizing the static spin-orbit rotation through the magnetic field orientation~\cite{Jock2018} is best to avoid inducing errors on the electron pair.  Ruling out static spin-orbit to control $c_1-c_2$, one could either shuttle electrons to micro- or nano-magnetic regions, or use a velocity-dependent spin-orbit effect (using the fact that a charged particle that has a component of its velocity orthogonal to an electric field will experience an effective magnetic field).  Alternatively, a network of additional nuclear spin qubits as reference could be maintained through pairwise parity checks and used to measure individual qubits with a pairwise parity check against a reference.

In assuming adiabatic shuttling during EPI, the electron polarization should be referenced to the instantaneous eigenbases along the respective paths.  In this sense, magnetic field direction need not be completely homogeneous along paths as long as adiabaticity with respect to spin states is maintained.  This will be aided by an energy gap due to Zeeman splitting given a modest magnetic field.  This gap should be large relative to the rate of change of the eigenstates which depends upon shuttle velocity.  Adiabaticity with respect to quantum dot orbital states is also important to ensure spin interactions are well-controlled.  With these caveats, $\phi(t)$ is actually determined by the difference in energy gaps of the instantaneous eigenstates, integrated over time, of the two paths.  These energy gaps may account for all magnetic-like interactions, including spin-orbit as well as hyperfine.   References to magnetic field gradients above are more precisely stated as energy gap differences.  Adiabatic shuttling on/off nuclei is also required and is easiest for isoelectronic species such as $^{29}$Si~\cite{Hensen2020-bf}, $^{117}$Sn, or $^{119}$Sn~\cite{Witzel2022} and is aided in a modest magnetic field by the gyromagnetic ratios of these nuclei being roughly three orders of magnitude smaller than an electron in silicon.  A magnetic field of one millitelsa is expected to be large enough to suppress flip-flop probabilities with a $^{119}$Sn to less than 2E-4 regardless of shuttling speed~\cite{Witzel2022}.

Most importantly, the only type of ancilla qubit error that can spread to corrupt data qubits during EPI is an electron spin flip.  Electron phase errors commute through the operation until the final measurement.  
Electron spin flips are suppressed during shuttling by energy gaps given sufficient magnetic field, tunnel coupling, and valley splitting.
Coherent electron shuttling over an effective distance of 10~$\mu$m in under 200~ns with an average fidelity of 99.5\% has been demonstrated in an isotopically purified Si/SiGe heterostructure device~\cite{DeSmet2025}.
EPI preserves coherence of the data qubits in a non-demolition manner while measuring their parity when spin flips are suppressed and the time-integrated hyperfine interaction for the hyperfine CPhase operation is precise. Measurement outcome errors can be suppressed exponentially through repeated parity measurements (e.g., by taking a majority vote) while the damage to data qubits from interactions with each fresh electron pair scales linearly. 
Our shuttling-based EPI protocol provides a noise-resilient and scalable approach to quantum error correction while maintaining the benefits of nuclear spin qubits.

\section{Quantum Error Correction with EPI}
\label{sec:NMR_and_DD}

As discussed above, EPI provides a robust means to perform parity measurements on collections of nuclei in the $Z$-basis. QEC of CSS codes requires parity check in two orthogonal bases.  We can accomplish this by applying global NMR operations to rotate the nuclear spins. A $Y_{\pi/2}$ NMR gate on all nuclei simultaneously effectively switches between $Z$ and $X$ bases (similar to a Hadamard gate in this respect). A cycle of parity checks then consists of a $Z$ parity check through $n$ measurement rounds of EPI, a global $Y_{\pi/2}$ NMR gate, an $X$ parity check through $n$ rounds of EPI, and another global $Y_{\pi/2}$ gate to return to the $Z$ basis.

\begin{figure}
\includegraphics[width=3.3in]{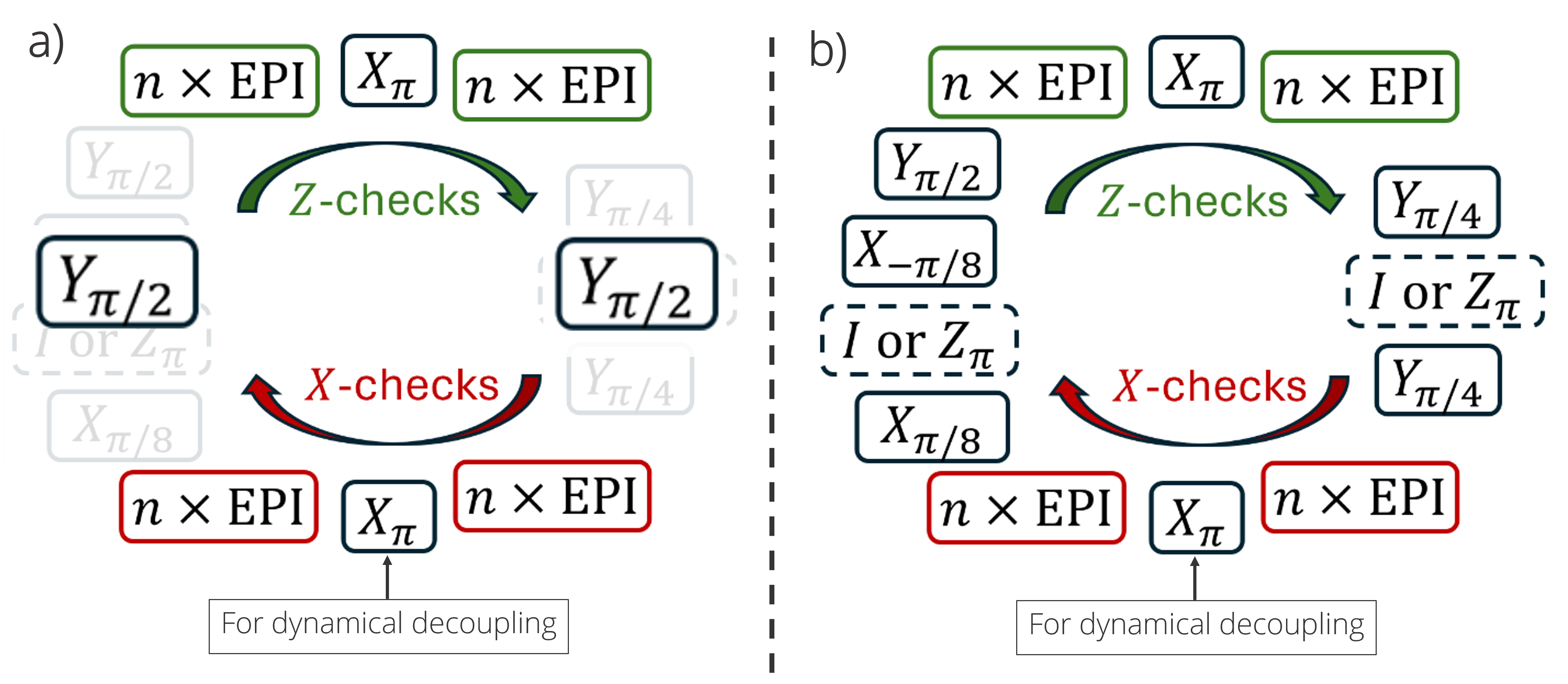}
\caption{
\label{fig:NMRcycle}
Global NMR cycle for universal quantum computation composed of NMR pulses (black boxes), EPI parity measurements (colored boxes), and selective $Z_{\pi}$ operations (dashed boxes). a) highlights the basic, default cycle for syndrome extraction; b) details the full cycle that enables universality with an effective $T$ (Hadamard) gate by applying $Z_{\pi}$ on the left (right) portion of the cycle.}
\end{figure}

Dynamical decoupling of the nuclear spins naturally fits into this cycle. The insertion of a global $X_{\pi}$ NMR gate in-between the $Y_{\pi/2}$ pulses provides DD for phase errors from $Y_{\pi/2}$ pulse miscalibration (e.g., from $B$-field inhomogenity). Further, repeating $n$ rounds of EPI before and after the $X_{\pi}$ gate implements DD of nuclear spin dephasing accrued during EPI [e.g., $b_1$ versus $b_2$ in Eq.~(\ref{eq:2parity})]. The global NMR cycle, shown in Fig.~\ref{fig:NMRcycle}(a), achieves QEC syndrome detection for logical idle of CSS codes.

Non-Clifford operations for universality require selective rotations on individual nuclei. Using global NMR for $X$ and $Y$ operations rotates all nuclear spin qubits. However, in the presence of a modest $B$-field ($\gtrapprox 1$~mT) a selective $Z$ operation can be accomplished through the interaction of an electron with a target nuclei. Recall that a hyperfine-CPhase gate induces a $Z_{\pm \pi/2}$ rotation on a nuclear qubit depending on the electron polarization.  Doubling the interaction time will induce a $Z_{\pm \pi} \cong Z_{\pi}$ rotation for an arbitrarily polarized electron (equivalent up to an irrelevant global phase).  Any $X$ or $Y$ rotation of a chosen subset of nuclear qubits can then be implemented using a combination of global NMR and selective $Z_{\pi}$ since $Z_{\pi} X_{\theta/2} Z_{\pi} X_{\theta/2} \cong X_{-\theta/2} X_{\theta/2} \cong I$ while $X_{\theta/2} X_{\theta/2} = X_{\theta}$ (and likewise replacing $X$ with $Y$) using $\cong$ to denote equality up to an irrelevant global phase.

The global NMR cycle depicted in Fig.~\ref{fig:NMRcycle}(b) is designed to achieve universal quantum computation. The boxes with ``$I$ or $Z_{\pi}$" depict local, selective $Z_{\pi}$ gates that may be induced by doubled hyperfine-CPhase gates as described above.  By choosing $I$ on the left and right of this cycle, Fig.~\ref{fig:NMRcycle}(b) reduces to Fig.~\ref{fig:NMRcycle}(a).
Choosing $Z_{\pi}$ for select qubits on the right of the cycle instead, the effect is $Y_{\pi/4} Z_{\pi} Y_{\pi/4} \cong Z_{\pi} (Z_{\pi} Y_{\pi/4} Z_{\pi}) Y_{\pi/4} \cong Z_{\pi} Y_{-\pi/4} Y_{\pi/4} \cong Z_{\pi}$ instead of $Y_{\pi/2}$ which has the same relative effect as a Hadamard gate with respect to the basis upon which EPI acts (those select qubits stay in the $Z$ basis rather than switching to the $X$ basis).  Choosing $Z_{\pi}$ on the left of the cycle, the overall effect when coming back to the $Z$-basis on the top of the cycle is $Y_{\pi/2} X_{-\pi/8} Z_{\pi} X_{\pi/8} Y_{\pi/2} \cong 
Z_{\pi} (Z_{\pi} Y_{\pi/2} Z_{\pi}) (Z_{\pi} X_{-\pi/8} Z_{\pi}) X_{\pi/8} Y_{\pi/2} 
\cong Z_{\pi} Z_{\pi/4}$ which is equivalent to a $T = Z_{\pi/4}$ gate up to $Z_{\pi}$.  Accounting for the remnant $Z_{\pi}$ in both cases is trivial with Pauli frame tracking that is standard practice in QEC. 


The global NMR schedule with EPI provides versatility to perform parity checks among different sets of nuclear spin qubits in two orthogonal bases.
This is sufficient for lattice surgery~\cite{Horsman2012} in which the choice of parity checks can be used to deform, merge, and split patches of CSS codes and effect logical parity measurements.  
Transversal CNOT operations can also be implemented with the aid of ancillary nuclear spin qubits to generate a CNOT operation from parity measurements.
We note that the locality of the code is only constrained by the distance of coherent electron shuttling opening up the possibility of using nonlocal quantum LDPC codes~\cite{Tillich2014, Breuckmann2021, Panteleev2022, Tremblay2022, Bravyi2024, Chadwick2026preprint}.
Our scheme with EPI and global NMR provides a comprehensive QEC implementation for CSS codes.

\section{Performance estimates}

We now estimate the performance of our EPI protocol with global NMR.  We will first discuss the potential fidelity of the hyperfine-CPhase gate and it's impact on EPI.  We then consider the effects of magnetic field inhomogeneity on global NMR in the context of our protocol which alternates between NMR-induced qubit rotations and EPI parity check measurements.

The hyperfine-CPhase gate has the potential for very good performance with spin flips (e.g., the $X \otimes X$ flip-flop error) suppressed below $10^{-5}$ with an applied field above a few millitesla and correlated $Z \otimes Z$ errors below $10^{-4}$ or much better when the hyperfine interaction is maximized~\cite{Witzel2022}.  However, significant isotopic enrichment is required to adequately reduce the electron phase error ($Z \otimes I)$. The phase error for a hyperfine CPhase gate time of $T$ is equal to $\left(1 - \exp{\left(-\left(T/T_2^*\right)^2\right)}\right)/2$. For Sn qubits in silicon, the phase error with $T_2^* \sim 10~\mu$s is estimated to be 1-10\%  and with $T_2^* \sim 100~\mu$s it is estimated to be 0.01-0.1\%, considering a range of quantum dot shapes~\cite{Witzel2022}. As isotopic enrichment increases, the variance of $T_2^*$ across quantum dots increases since $T_2^*$ becomes a strong function of the individual locations of $^{29}$Si atoms in the lattice. As an example, at 50~ppm $^{29}$Si, over 90\% of such considered quantum dots will have $T_2^* > 10~\mu$s~\cite{Witzel2022}; at 2~ppm $^{29}$Si, we found that over 85\% will have $T_2^* > 100~\mu$s under the same considerations. Shuttling provides a means to select the best performing sites while quickly moving through undesirable sites.
Electron phase errors can potentially be mitigated using feedback control~\cite{Stuyck2024, Steinacker2025, WitzelNucMagNoise} if calibrated singlet/triplet rotations are feasible (via aforementioned micro- or nano-magnetic regions or spin-orbit effects). As previously mentioned, electron phase errors do \emph{not} propagate to the nuclear qubits and contribute solely to the parity measurement outcome error, which can be overcome with sufficient repetition of rounds of EPI.

Nuclear spin qubit error induced by magnetic field inhomogeneity, both in the static (DC) $B_0$ as well as the oscillating (AC) NMR $B_1$ fields, is also an important consideration.
To analyze the impact of magnetic field inhomogeneity on QEC performance, we deduce an effective detector error model (DEM)~\cite{DEM}, accounting for error probabilities with various combinations of outcome effects from subsequent parity measurements due to imperfections in the two kinds of net NMR rotations, $X_{\pi}$ and $Y_{\pi/2}$, used in our syndrome extraction protocol and interspersed with EPI parity measurements for logical idle (e.g., following the protocol of Fig.~\ref{fig:NMRcycle}(a)).

Our analysis below treats the static (perfectly correlated in time) inhomogeneity of magnetic fields that will be increasingly challenging to minimize as devices scales.  Spatial correlations may also be strong. For example, it might be reasonable to assume that the length scale of magnetic inhomogeneity will be long compared with the spatial extent of logical patches but short over the scale of a system of many logical patches.  Therefore, the constituents of many logical patches may be similarly shifted away from an ideal target set for global NMR. In this case, the effects of spatial correlations may add constructively (worst-case) or subtract destructively (best-case) depending upon the codespace of a patch.  There is a simple no-cost solution to the danger of the worst-case.  By allowing the codespace to wander via Pauli frame tracking rather than active error correction, it has been shown~\cite{WitzelGoWithFlow2026} that spatial correlations of this nature can be disregarded when codes are larger than distance $3$.
With this justification, we model rotation errors due to magnetic field inhomogeneity as an independent noise source for each nuclear spin qubit individually.

\begin{figure}
\includegraphics[width=3.3in]{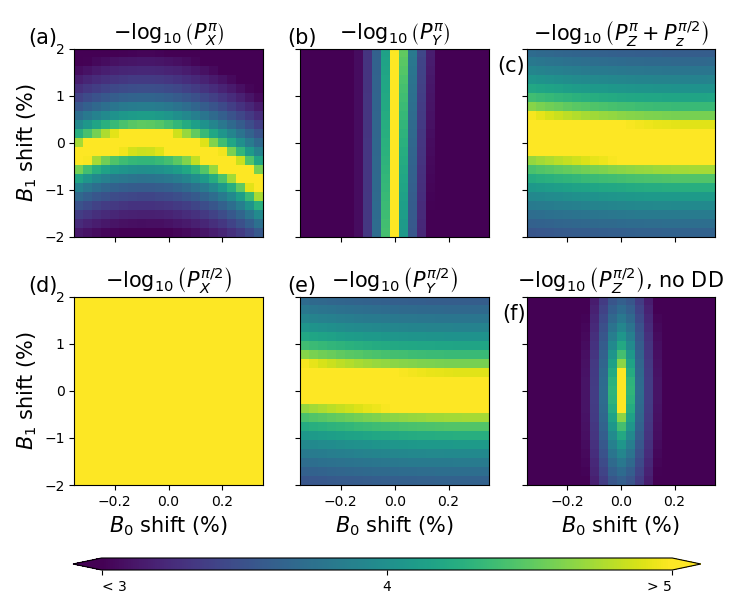}
\caption{
\label{fig:NMR_DEM_1mT_100uT}
DEM for $B_0 = 1$~mT, $B_1 \approx 100~\mu$T both with DD (a)-(e) and without it (f).  With DD, frequencies, durations, and $B_1$ were $15.93$ ($15.95$)~kHz, $313.6$ ($658.1$)~$\mu$s, and $99.9$ ($95.2$)~$\mu$T
for $Y_{\pi/2}$ ($X_{\pi}$).  Without DD, $B_0$ ($B_1$) were $996$~$\mu$T ($99.5$~$\mu$T) with $15.89$~kHz frequency and $314.6$~$\mu$s duration.
}
\end{figure}v

\begin{figure}
\includegraphics[width=3.3in]{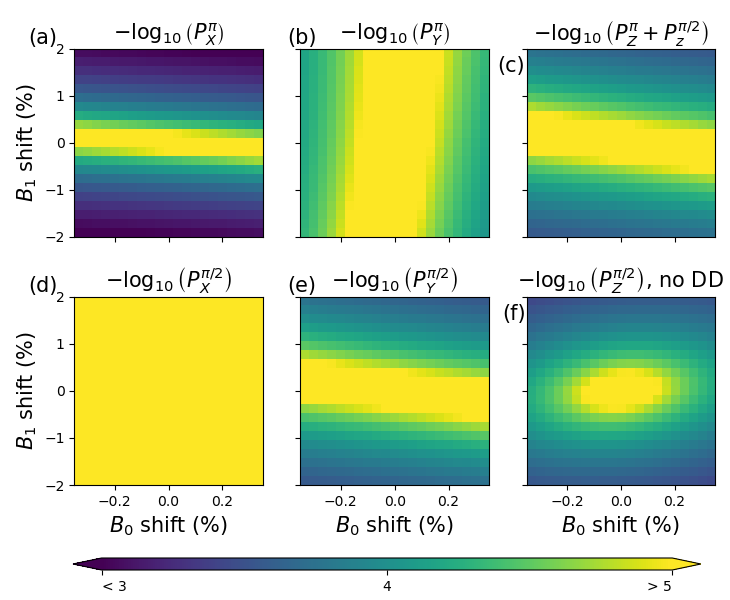}
\caption{
\label{fig:NMR_DEM_1mT_1ishmT}
DEM for $B_0 = 1$~mT, $B_1 \approx 1~$mT both with DD (a)-(e) and without it (f).  With DD, frequencies, durations, and $B_1$ were $16.02$ ($15.00$)~kHz, $31.20$ ($66.68$)~$\mu$s, and $99.9$ ($95.2$)~$\mu$T
for $Y_{\pi/2}$ ($X_{\pi}$).  Without DD, $B_0$ ($B_1$) were $1.064$~mT ($984$~$\mu$T) with $15.96$~kHz frequency and $31.32$~$\mu$s duration.
}
\end{figure}

Our DEM ascribes a probability to $X$, $Y$, and $Z$ errors following each NMR gate for a given nuclear qubit depending upon its local, static DC and AC magnetic field offsets.  The $X$, $Y$, and $Z$ errors each correspond to different combinations of effects on subsequent parity measurements (e.g., $X$ induces a parity flip in the same basis, $Z$ induces a parity flip in the opposite basis effected by the next $Y_{\pi/2}$ rotation, and $Y$ induces a parity flip in both bases).
In our notation, $P^{G}_{E}$, we use a superscript to indicate the gate $G$ as $\pi$ or $\pi/2$ and subscript to indicate the error $E$ as $X$, $Y$, or $Z$.
Figure~\ref{fig:NMR_DEM_1mT_100uT} and Figure~\ref{fig:NMR_DEM_1mT_1ishmT} show our DEM findings with $B_1 \approx 100 \mu$T and $1$~mT respectively, assuming $B_0 = 1~$mT, ranging over failure probabilities between $10^{-3}$ (blue) and $10^{-5}$ (yellow).
There is no distinction between ascribing $Z$ error probability following a $Y_{\pi/2}$ gate versus a $Z$ error probability following the subsequent $X_{\pi}$ gate since both have the same impact on parity measurements, so we only report the sum of these probabilities.  To understand the impact of DD, these figures also show the $Z$ error of the $Y_{\pi/2}$ gate without DD (no $X_{\pi}$ gates).  We only show the $Z$ error because DD has no significant impact on $P_X^{\pi/2}$ or $P_Y^{\pi/2}$ (due to the interspersed parity measurement projections).  By comparing panels (c) and (f), we can see the benefit that DD provides with respect to correcting phase errors of the $Y_{\pi/2}$ gate. However, introducing the $X_{\pi}$ gate to effect DD contributes additional noise that appear, from these figures, to essentially counteract these benefits.  On the other hand, DD can cancel nuclear spin phase errors that build up in between NMR pulses that may be important.
Also, $X_{\pi}$ could be implemented more robustly using adiabatic-based approaches~\cite{GuryOdelin2019}.

We invoke the  Choi-Jamiołkowski isomorphism to arrive at our DEMs and produce Figs.~\ref{fig:NMR_DEM_1mT_100uT} and~\ref{fig:NMR_DEM_1mT_1ishmT}.  Specifically, we simulate an experiment in which parity measurements are repeated on a pair of nuclear qubits initialized in a Bell state, one qubit experiencing the error from $B$-field offsets and the other qubit experiencing no error.  This thought experiment captures both the effects of projective parity measurements as well as non-projected, coherent phase errors.  The latter could not be captured by alternating NMR and single-qubit measurements, but is fully captured in the two qubit state space (e.g., $\lvert \Uparrow \Downarrow \rangle + e^{i \alpha} \lvert \Downarrow \Uparrow \rangle$ for an odd parity projection or $\lvert \Uparrow \Uparrow \rangle + e^{i \alpha} \lvert \Downarrow \Downarrow \rangle$ for an even parity projection).  Results are based on single-frequency NMR without using the rotating wave approximation where we have determined the frequency and duration of an ideal NMR pulse and adjusted $B_1$ values about targets using Nelder-Mead optimization.
From this simulation, we determine probabilities, $P_{abc}$, for different detector combinations for ideal parity measurements following each NMR gate of a $Y_{\pi/2}$-$X_{\pi}$-$Y_{\pi/2}$ sequence.  More specifically, the simulation executes ${\mathbb P}^c \left(\tilde{Y}_{\pi/2} \otimes Y_{\pi/2}\right) {\mathbb P}^{b \oplus a} \left(\tilde{X}_{\pi} \otimes X_{\pi}\right) {\mathbb P}^a \left(\tilde{Y}_{\pi/2} \otimes Y_{\pi/2}\right)$ as an operator acting from right to left where ${\mathbb P}^x$ are parity projection measurements whose outcome is assigned to $x$ ($b \oplus a$ is the outcome relative to the first, rightmost, projector)
and $\tilde{Y}_{\pi/2}$ and $\tilde{X}_{\pi}$ are the non-ideal NMR gates; that is, $a=1$, $b=1$, or $c=1$ respectively represent a relative change of parity in the intended basis of each parity measurement ($b$ is relative to $a$ since there is no change of basis between their associated projections).
This is sufficient to capture all parity changes of an indefinitely repeated cycle [Fig.~\ref{fig:NMRcycle}(a)] due to errors induced by the first $Y_{\pi/2}$ gate or the $X_{\pi}$ gate since both of these are proceeded by both $Z$ and $X$ parity measurements.
Since a DEM would dictate $P_{abc}$ values, we invert this relationship to arrive at a DEM from calculated $P_{abc}$ probabilities.  To a first order approximation, we have $P_{100} \approx P_x^{\pi/2}$, $P_{010} \approx P_x^{\pi}$, $P_{101} \approx P_y^{\pi/2}$, $P_{011} \approx P_y^{\pi}$, and $P_{001} \approx P_z^{\pi/2} + P_z^{\pi} + P^{\pi/2}_x$.  Therefore, $P_z^{\pi/2} + P_z^{\pi} = P_{001} - P_{100}$, and we have determined all of our DEM parameters from $P_{abc}$.  The $P_{110} \approx P_x^{\pi/2} P_x^{\pi} + P_y^{\pi/2} P_y^{\pi}$ and $P_{111} \approx P_x^{\pi/2} P_y^{\pi} + P_y^{\pi/2} P_x^{\pi}$ probabilities are used as sanity checks to confirm the applicability of our first-order approximation; their discrepancies are indeed negligibly small.

From Fig.~\ref{fig:NMR_DEM_1mT_100uT}, we can conclude that, at an operating point of $B_0 = 1~$mT and $B_1 \approx 100~\mu$T, high fidelity operations below 1E-3 may be achieved with a $B_0$ uniformity of about 0.1\% and a $B_1$ uniformity of about 1\%.  These inhomogeneity requirements are relaxed by increasing $B_1$ due to NMR power broadening as seen in Fig.~\ref{fig:NMR_DEM_1mT_1ishmT}.  Reducing $B_0$ brings two advantages with respect to NMR considerations, it reduces NMR power by reducing the resonance frequency proportionally with $B_0$ and it improves the power broadening which depends upon $B_1$ relative to $B_0$.
On the other hand, a small $B_0$ can make adiabaticity during EPI more challenging. 

\section{Conclusion}

We have presented the EPI technique to measure the parity of a collection of nuclear spins in distinct quantum dots by shuttling a pair of electrons such that each nuclear spin qubit interacts with one of these electrons via a noise-resilient hyperfine-CPhase gate~\cite{Witzel2022}.  Provided high-fidelity singlet-state preparation of the electron spin pair and well-preserved electron polarization, the noise-resilience of this CPhase gate should translate to excellent nuclear spin coherence preservation during this parity measurement.  This is further enhanced with refocusing pulses between parity measurements as an efficient implementation of DD.  Electron dephasing errors incurred during shuttling due to spin-orbit interactions will only impact the measurement outcome fidelity without impacting the nuclear spin coherence.  Given DD refocusing and long intrinsic nuclear spin lifetimes, EPI parity measurements may be repeated to  improve measurement outcome certainty exponentially while only increasing the nuclear spin error probability linearly.

Scalable QEC of nuclear spin qubits, in the context of any CSS code, is enabled by our shuttle-based EPI parity measurement operation combined with global NMR.  Global NMR can be used to switch bases to effect both $X$ and $Z$ parity checks and perform DD.  Selective hyperfine-induced $Z_{\pi}$ gates, via interacting a nuclear spin qubit with an arbitrarily-polarized electron spin, complete a universal gate set.  Finally, we present theoretical DEM probabilities under the consideration of static uncertainties in the $B_0$ and $B_1$ of the global NMR controls.

We would like to acknowledge valuable discussions and insights with Andrew Landahl, Malcolm Carroll, Rick Muller, Nathan Bishop, Robin Blume-Kohout, John Gamble, Anand
Ganti, N. Tobias Jacobson, Ines Montano, Erik Nielsen, and Kevin Young.
Research was sponsored by the Army Research Office and was accomplished under
Cooperative Agreement Number W911NF-22-2-0037. The views and conclusions contained in this document are those of the authors and should not be interpreted as representing the official policies, either expressed or implied, of the Army Research Office or the U.S. Government. The U.S. Government is authorized to reproduce and distribute reprints for Government purposes notwithstanding any copyright notation herein.
Sandia National Laboratories is a multimission laboratory managed and operated by National Technology \& Engineering Solutions of Sandia, LLC, a wholly owned subsidiary of Honeywell International Inc., for the U.S. Department of Energy's National Nuclear Security Administration under contract DE-NA0003525.  This paper describes objective technical results and analysis. Any subjective views or opinions that might be expressed in the paper do not necessarily represent the views of the U.S. Department of Energy or the United States Government.

\bibliography{bibliography}
\end{document}